\documentclass[3p,times,procedia]{elsarticle}
\usepackage{nupha_ecrc}
\usepackage{here}

\volume{00}

\firstpage{1}

\journalname{Nuclear Physics A}

\runauth{}

\jid{nupha}

\jnltitlelogo{Nuclear Physics A}

\usepackage{amssymb}
\biboptions{sort}

\usepackage[figuresright]{rotating}

\begin{document}

\begin{frontmatter}

%% Title, authors and addresses

%% use the tnoteref command within \title for footnotes;
%% use the tnotetext command for the associated footnote;
%% use the fnref command within \author or \address for footnotes;
%% use the fntext command for the associated footnote;
%% use the corref command within \author for corresponding author footnotes;
%% use the cortext command for the associated footnote;
%% use the ead command for the email address,
%% and the form \ead[url] for the home page:
%%
%% \title{Title\tnoteref{label1}}
%% \tnotetext[label1]{}
%% \author{Name\corref{cor1}\fnref{label2}}
%% \ead{email address}
%% \ead[url]{home page}
%% \fntext[label2]{}
%% \cortext[cor1]{}
%% \address{Address\fnref{label3}}
%% \fntext[label3]{}

%% Instructions from Editor: Please use the following \dochead only in the preprint version (e-print arXiv etc.); 
%% use empty \dochead{} when submitting to Nuclear Physics A!
\dochead{XXVIIth International Conference on Ultrarelativistic Nucleus-Nucleus Collisions\\ (Quark Matter 2018)}
%\dochead{}
%% Use \dochead if there is an article header, e.g. \dochead{Short communication}
%% \dochead can also be used to include a conference title, if directed by the editors
%% e.g. \dochead{17th International Conference on Dynamical Processes in Excited States of Solids}

\title{Open Heavy Flavors in Nuclear Collisions: Theory Overview}

%% use optional labels to link authors explicitly to addresses:
%% \author[label1,label2]{<author name>}
%% \address[label1]{<address>}
%% \address[label2]{<address>}

\author{Pol B Gossiaux}

\address{SUBATECH, UMR Université de Nantes, IN2P3/CNRS, IMT Atlantique; Nantes, France}

\begin{abstract}
I review the current status and some prospects of theoretical studies on open heavy flavor physics in nuclear collisions at RHIC and LHC energies.
\end{abstract}

\begin{keyword}
%% keywords here, in the form: keyword \sep keyword
heavy quarks, quark gluon plasma, ultrarelativistic heavy ion collisions

%% MSC codes here, in the form: \MSC code \sep code
%% or \MSC[2008] code \sep code (2000 is the default)

\end{keyword}

\end{frontmatter}

%%
%% Start line numbering here if you want
%%
% \linenumbers

%% main text
\section{Introduction and structuration}
\label{}

Production of heavy flavors (HF) in ultrarelativistic heavy ion collisions is usually advocated as an ideal probe of the deconfined phase, i.e. the quark gluon plasma created in those collisions. Several facts contribute to this statement: Heavy quarks (HQ) are dominantly produced during the initial nucleonic collisions and are then conserved through the time evolution (even at LHC where thermal production is still a limited fraction); they are then strongly affected by the QGP phase and much less by the ensuing hadronic phase, thanks to their large mass $m_Q$. This large scale also helps in designing pQCD calculations for their initial production as well as developing some simplified physical picture of their interaction with the QGP based on the notion of a relaxation time $\tau_{\rm relax}\propto \frac{m_Q}{T^2}$, which follows a clear hierarchy from $s$ to $c$ to $b$ quarks. Some schemes of energy loss calculation also take explicit advantage of the large mass assumption (see f.i. \cite{Liu:2017qah}). 

Since large quenching of those HF has been observed both at RHIC and at LHC (see \cite{Andronic:2015wma} for a recent review), it is desirable to achieve a quantitative understanding of the experimental results that can also be expressed in terms
of intelligible quantities, namely the so-called Fokker-Planck (FP) coefficients. Given one HQ propagating in the hot medium and undergoing elastic interactions with its constituents, one can indeed -- adopting a classical picture -- describe its trajectory by the means of stochastic equations whose first and second moments satisfy the following laws:
\begin{equation}
-\frac{{\rm d}}{{\rm d}t}\langle \vec{p} \rangle=-\vec{A}(\langle \vec{p} \rangle,T)  = -\eta_D(\vec{p},T)  \langle \vec{p} \rangle
\quad {\rm and} \quad
\frac{{\rm d}}{{\rm d}t}\langle \vec{p}_{T,i} \vec{p}_{T,j} \rangle= \kappa_T(\langle \vec{p} \rangle,T)\,\delta_{i,j}
\label{eq:diffusion}
\end{equation}
where $\eta_D [{\rm fm}^{-1}]$ can be interpreted as the inverse relaxation time while $\kappa_T [{\rm GeV}^2{\rm fm}^{-1}]$ is the transverse diffusion coefficient, directly proportional to the transport coefficient $\hat{q}$ ($\hat{q}=2\kappa_T$). Similar relation holds along the longitudinal direction and implies the longitudinal diffusion coefficient $\kappa_L$. Equations (\ref{eq:diffusion}) are generically valid whatever the precise effective degrees of freedom (dof) scattering with HQ. If the HQ undergoes radiative scatterings, the energy loss pattern becomes more complex and coherence effects may play an important role. In this later case, the total momentum loss over the path length $L$ acquires a contribution $\left(\Delta \vec{p}\right)_{\rm rad}$ $\propto L^\alpha$ with $\alpha\ge 2$. However, in most of the existing schemes, this contribution can still be related to the FP coefficients, what confirms their special importance for describing and understanding the physics of HQ interacting with QGP and the need to be able to evaluate them in a way or another. While those coefficients are in general not related to each other, a noticeable exception is found at small momentum $p\sim m_Q$ (and not too large temperature) -- where the diffusive Langevin regime applies and where the 3 coefficients are related by a generalized Einstein relation.  
% cite some review.
For $p\rightarrow 0$, one has $\kappa_L=\kappa_T=\kappa$ and this relation even simplifies to $\kappa(p=0)=2 T m_Q \eta_D(p=0)$. For historical reasons, one prefers to write those coefficients in terms of the spatial diffusion coefficient $D_s$, which is thus a good gauge of the HQ coupling with the QGP:
\begin{equation}
{(2\pi T) D_s} =\frac{4 \pi  T^3}{{\kappa}}= 
\frac{2 \pi  T^2}{E_Q  {\eta_D}} \quad \Rightarrow \quad \tau_{\rm relax}={\eta_D^{-1}}= {(2\pi T) D_s} \times \frac{m_Q}{2 \pi  T^2}.
\end{equation}
It is precisely in this low momentum regime that the $D_s$ can be evaluated from first principle resorting to lQCD simulations, still affected by large uncertainties as 
% LQCD refs 
values of the order of $6 \pm 2$ have been obtained for $T$ around $T_c$, leading to $\tau_{\rm relax}\approx (3 \pm 1.5)\,{\rm fm}$. Beyond this regime it is often advocated that pQCD calculations of the energy loss are able to reproduce the quenching observed in the experimental data down to $p_T\approx 10-20\,{\rm GeV}/c$. 
% cite Djordjevic. 
In the intermediate $p_T$ regime, where lies the bulk of the data, one has however to rely on the numerous effective models (see table \ref{models}) that were developed and calibrated over the 10 past years and which are used as a means to make the junction\footnote{ As most of those models are not exclusively deduced from the fundamental theory, it is fair to say that we are for the time mainly focused on gauging and understanding the HQ interaction with the QGP (more than really probing this medium extensively with HQ).} between the experimental data and the fundamental quantities such as the FP coefficients. 
% cite a couple of them. 
Nowadays, one of the burning questions for the field is to know whether we are collectively in the position to extract those coefficients with a decent precision (the desired level of accuracy being a subtle issue in itself) and a reasonable consensus. In this respect, it is crucial to notice that each "model", besides its core ingredients (HQ - QGP basic interaction and its transport implementation), also relies on extra ingredients such as the hadronization prescription, the bulk description, etc. that can lead to important deviations in the final yield of HF hadrons. In these proceedings, I summarize recent progresses made in this direction and provide some prospects for the near future. Further references and discussions can be found in recent reviews such as
\cite{Aarts:2016hap,Andronic:2015wma,Prino:2016cni,Rapp:2018qla}.

\begin{table}
\begin{tabular}{|c|c|c|c|c|}
\hline
  & elastic & elastic + radiative & radiative & other \\
\hline
transport coeff. based & TAMU &   Duke & ASW & AdS/CFT\\
(Langevin,\ldots) & Catania LV & & & POWLANG lQCD
 \\ 
 &  POWLANG HTL & &  & DABMOD \\
  &   & &  & S. Li et al \\
\hline
cross section or $|\mathcal{M}|^2$ based & {\em AMPT} & Djordjevic et al & ${\rm SCET_{G,M}}$ & \\
(Boltzmann,\ldots) & MC@sHQ el & MC@sHQ el + rad & & \\
& {\em URQMD} & {\em BAMPS} & &  \\
& {\em PHSD} & CUJET3  & &  \\
& {\em Catania BM}  & LBL-CCNU & &  \\
& & {VNI/BMS} & &  \\
% & & Abir \& Mustafa & &  \\
& & HYDJET++& &  \\
& & LIDO & &  \\
\hline
\end{tabular}	
\caption{Classification of a large variety of models aiming at describing HF production in URHIC; in italic, those including a full transport implementation both in the light and in the heavy sector (LIDO: Poster by W. Ke at this conference).}
\label{models}
\end{table}
% todo provide a table of models 

\section{Lessons from $R_{AA}$ and $v_2$ of D mesons}
The nuclear modification factor $R_{AA}$ and elliptic flow $v_2$ of D mesons are considered as the ground observables which provide basic constrains on the models and the transport coefficients. The effects ruling the global pattern of these observables are by now rather well established, at least at a qualitative level: The large depletion observed for the $R_{AA}$ at high $p_T$ (as well as its recovery towards unity at even larger $p_T$) is primarily understood as due to HQ radiative energy loss in the QGP, while at low $p_T$, HQ -- especially c quarks -- achieve a high degree of equilibration with the QGP and thus benefit from its local flow, what can possibly lead to a so-called "flow bump" in the $R_{AA}$, whose magnitude however strongly depends on other ingredients like initial state effects (shadowing) or the hadronization of HQ. At intermediate $p_T$, the energy loss mechanisms become more complicated: The Langevin picture, valid at low $p_T$, is no longer correct while coherence plays an increasing role in the radiative energy loss; adopting a high $p_T$ perspective, the eikonal limit stops to apply and energy loss fluctuations need to be considered. This leads to more involved schemes \cite{Abir:2015hta} 
in which the transport coefficient $\hat q$ is no longer the unique parameter. The elliptic flow pattern is also traditionally understood as a gradual transition from collective effects at small $p_T$ (with the same ingredients as for the $R_{AA}$) to anisotropy of the path length in the energy loss at high $p_T$, while all these effects compete at intermediate $p_T$. Recently an alternate explanation -- the so called "escape mechanism" was however suggested as responsible for the $v_2$ in the light sector \cite{He:2015hfa}, where it could be understood as a refined core-corona picture, as well as for the HQ~\cite{Li:2018leh} (see discussion in section \ref{section_flow}).

Whereas qualitative reproduction of $R_{AA}$ and $v_{2}$ patterns can be achieved with nearly each model containing the aforementioned ingredients, quantitative agreement can only be obtained for specific energy loss models. In \cite{Andronic:2015wma}, systematic comparisons were made between a large class of models available at that time and HF $R_{AA}$ and $v_2$ results both at RHIC and LHC. Although most of the models could accommodate the experimental results within statistical and systematic uncertainties, it was realized that a lot of models predicted too small $v_2$ as compared to the data, especially the ones resorting to both collisional and radiative energy loss. Besides, the $D_s$ coefficient associated with some of the models found in quantitative agreement with the data have been shown to vary by a factor 5     
\cite{Prino:2016cni}, what is obviously not satisfactory. In a recent study \cite{Das:2015ana}, the Catania group has advocated that several ingredients (see fig.\ref{fig:2} right) could reduce the tension between the $R_{AA}$ and the $v_2$, as this flow develops until very late times (i.e. for $T$ of the order and even lower than the critical temperature $T_c$) while the $R_{AA}$ pattern is achieved pretty early: a) inclusion of hadronic rescatterings, b) hadronization through coalescence at low $p_T$ and c) $T$ dependence of the drag coefficient $\eta_D(T)$. Each of the two first ingredients can lead to $\approx 1\%$ increase of the $v_2$, while the last one can lead to 3\% increase when passing from pQCD energy loss 
($\eta_D \propto T^2$) to non perturbative models like Quasi Particle Model, pHSD or the T-matrix approach from TAMU with a strong potential for which $\eta_D \propto T^0$ thus leading to a stronger weight of $T\approx T_c$ for the same $R_{AA}$. It was also shown in \cite{Das:2015ana} that for the same generic HQ-medium interaction, a Boltzmann transport leads to a $\approx 1\%$ increase of the $v_2$ with respect to a Langevin transport, when both strengthes of the interaction are tuned in order to reproduce a realistic $R_{AA}$. This can be understood based on the reduction of the longitudinal fluctuations in the Langevin transport once the FDT is imposed, leading to a smaller coupling if both transports are adjusted to the same $R_{AA}$, hence a smaller $v_2$. Following \cite{Das:2015ana}, incorporating all these four ingredients should thus be considered as the natural way for the models to cope simultaneously with both the $R_{AA}$ and the $v_2$, as demonstrated in the past year by the Catania group within a more sophisticated approach -- presented at this conference -- relying on a full Boltzmann transport \cite{Scardina:2017ipo} whose cross sections in the light sector are tuned to reproduced a fixed $\eta/s(T)$ ratio, with ensuing $D_s$ coefficients found in the bulk of the lQCD data and a factor $\approx 5$ smaller than the pQCD calculation.

\begin{figure}[H]
	\centering
	\includegraphics[width=0.4\textwidth]{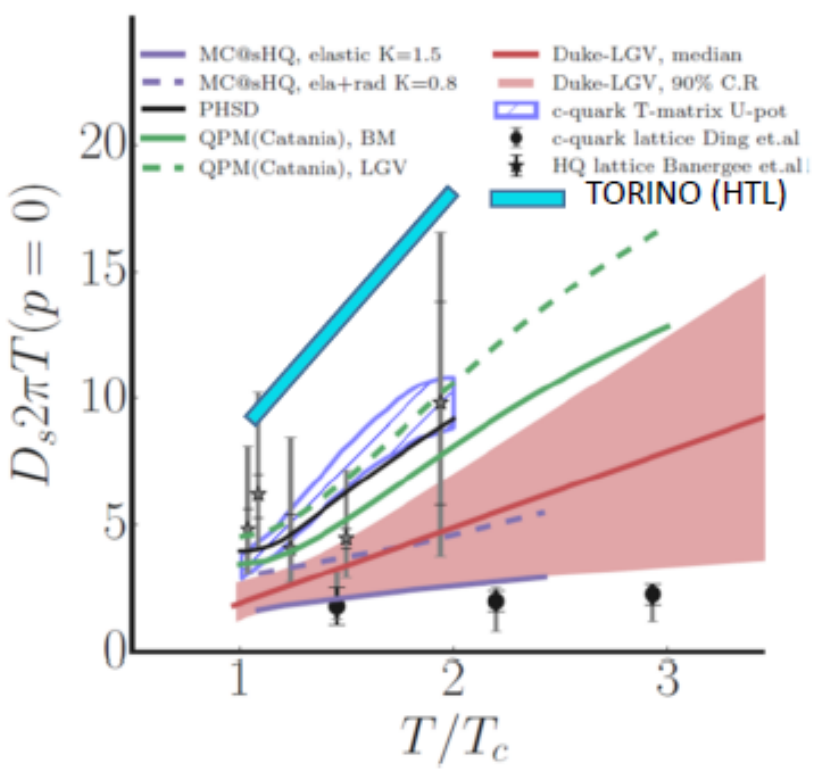} 
	\caption{Summary plot of various $D_s$ coefficient extracted from recent models compatible with $R_{AA}-v_2$ data (lines/bands) compared to lQCD data.}
	\label{fig:1}
\end{figure}

Another noticeable achievement this year is the application of state of the art Bayesian methods by the Duke group~\cite{Xu:2017obm} in order to perform data-driven extraction of the diffusion coefficient $D_s$ based on an extended set of experimental results. For this purpose, the authors have complemented the pQCD value of $D_s(T,p)$ by a non-perturbative part of tunable range in momentum space and of tunable slope as a function of T. The total $D_s$ coefficient then enters the energy loss computation, with a radiative component modeled through the higher twist approach. They conclude to some significant contribution from non perturbative effects up to $p_T\approx 20\,{\rm GeV}/c$, while the $T$ dependence of $D_s$ cannot be extracted precisely with this method, leaving room for future improvements. According to me, one of them -- sticking to the spirit of the method -- would be to allow some extra free parameter in the radiative component as f.i. the thermal gluon mass. In all cases, there has been over the past years an increasing evidence that effects going beyond LO pQCD should be taken into account around $T_c$ in order to be able to cope with experimental data at low and intermediate $p_T$ (see has well, \cite{Cao:2016gvr}), translating into rather small values of $D_s$, as illustrated on fig.~\ref{fig:1}. One should however refrain from drawing too strong conclusions from the inspection of $D_s$ alone, as the momentum dependence of the drag coefficient also plays an important role on $R_{AA}$ and $v_2$ and can differ quite strongly, as demonstrated in \cite{Prino:2016cni} by comparing POWLANG-HTL and TAMU models.     

Looking at the physics from a high-$p_T$ perspective, several sophisticated pQCD-based schemes (DGLV, higher twist, ${\rm SCET_{G,M}}$) have been developed over the last few years in order to properly take into account the coherence effects ruling the radiative energy loss as well as the role of the mass, with correct agreement for both the $R_{AA}$ of light and heavy mesons, sometimes leading to unexpected results such as the inversion of the mass hierarchy of jet quenching effects with prompt b-jet substructure \cite{Li:2017wwc}. However, not all the models have up to now considered a realistic medium evolution, that prevents precise quantitative conclusions. Progresses in this direction have been presented at this conference for the DGLV - DREENA-B approach~\cite{Zigic:2018ovr}, in which a Bjorken scenario was considered for the QGP evolution, leading to good success in explaining $R_{AA}$ and $v_2$ of both charged particles {\em and} HF mesons for all centralities and $p_T > 10\,{\rm GeV}$. Dealing with HQ produced at high $p_T$ naturally goes along with adopting a jet framework. In \cite{Cao:2017crw}, a multi-stage approach is implemented in which a medium-modified PYTHIA is first applied from a high virtuality scale $Q\approx p_T$ down to a low virtuality scale $Q_0\approx m_Q$, after what on-shell time-ordered propagation is performed according to the model of \cite{Cao:2015hia} with radiative component including finite energy corrections as calculated by \cite{Abir:2015hta}. D and B mesons $R_{AA}$ from CMS are well reproduced for $p_T> 10\,{\rm GeV}/c$, while finite energy corrections appear to have at most a 5\% influence for D-mesons. In a more ambitious treatment, the conversion of quenched gluons into c and b quarks in the jet evolution should be considered as well and this could lead to a reduction of the mass effect if those gluons have traveled significant distances in the QGP. For an exact treatment of such effects, mass dependent splitting functions have recently been calculated in the framework of a soft collinear effective theory \cite{Kang:2016ofv}. Despite all these achievements and good agreement between the theory calculations and experimental data, it should be noted that benchmarking results for basic quantities such as the total energy loss $\Delta E$ as a function of the path length $L$ do not systematically agree between various schemes, as recently demonstrated in \cite{Rapp:2018qla}, what obviously deserves some further work.  

\section{Harmonic flows}
\label{section_flow}
As recognized quite early, harmonic flows of the heavy mesons contain very rich information as regards the coupling of HQ with the QGP medium. For a while only $v_2$ of D mesons and non-photonic single electrons were accessible but accumulated statistics now allows to chase higher harmonics $v_n$, while precises measurements for $v_n(B)$ are expected in the next runs. In  \cite{Nahrgang:2013saa}, a systematic decrease of the $v_n$ over spatial eccentricity ratio $\epsilon_n$ was found for heavier mass, higher harmonics and more peripheral reactions, that was interpreted as due to the inertia of HQ, hence limiting they ability to couple efficiently with the medium. As $\epsilon_n$ cannot be measured directly in experiments, it was then suggested to proceed to so called "event shape engineering" methods to investigate the correlation between the flow of the HF mesons and light hadrons on an event-by-event basis. While recent investigations~\cite{Prado:2016szr,Gossiaux:2017zto} have concluded to a linear correlation between elliptic flows, a stronger correlation was found in the DABMOD model~\cite{Prado:2016szr} when a constant drag coefficient (vs $T$) was chosen, consistently with the late building of the harmonic flows for HQ. Similar conclusion was reached by the Catania group and presented at QM by S. Plumari who compared the correlations for two kinds of interactions (rescaled pQCD and Quasi-Particle model) and found an increasing difference for larger and larger flow harmonics. This type of study opens very promising perspectives as for the use of higher flow harmonics as a discriminating tool between models in the near future\ldots At this conference, a detailed analysis of flow building in POWLANG~\cite{Beraudo:2017gxw} was presented by A. Beraudo. It reveals that the final $v_2$ is a subtle interplay between positive and negative contributions accumulating along time, with a strong final positive peak. This scenario strongly contrasts with the monotonic behavior found in the context of the escape mechanism~\cite{Li:2018leh} where the $v_2$ is found to be built from smoothly decreasing positive contributions. I feel it would be pretty enlightening if each model could come with a similar study in order to be able to compare not only the flow magnitude but also the way it develops during the URHIC.

\section{Recent collective actions}
\label{collect}
\begin{figure}[H]
	\centering
	\includegraphics[width=0.4\textwidth]{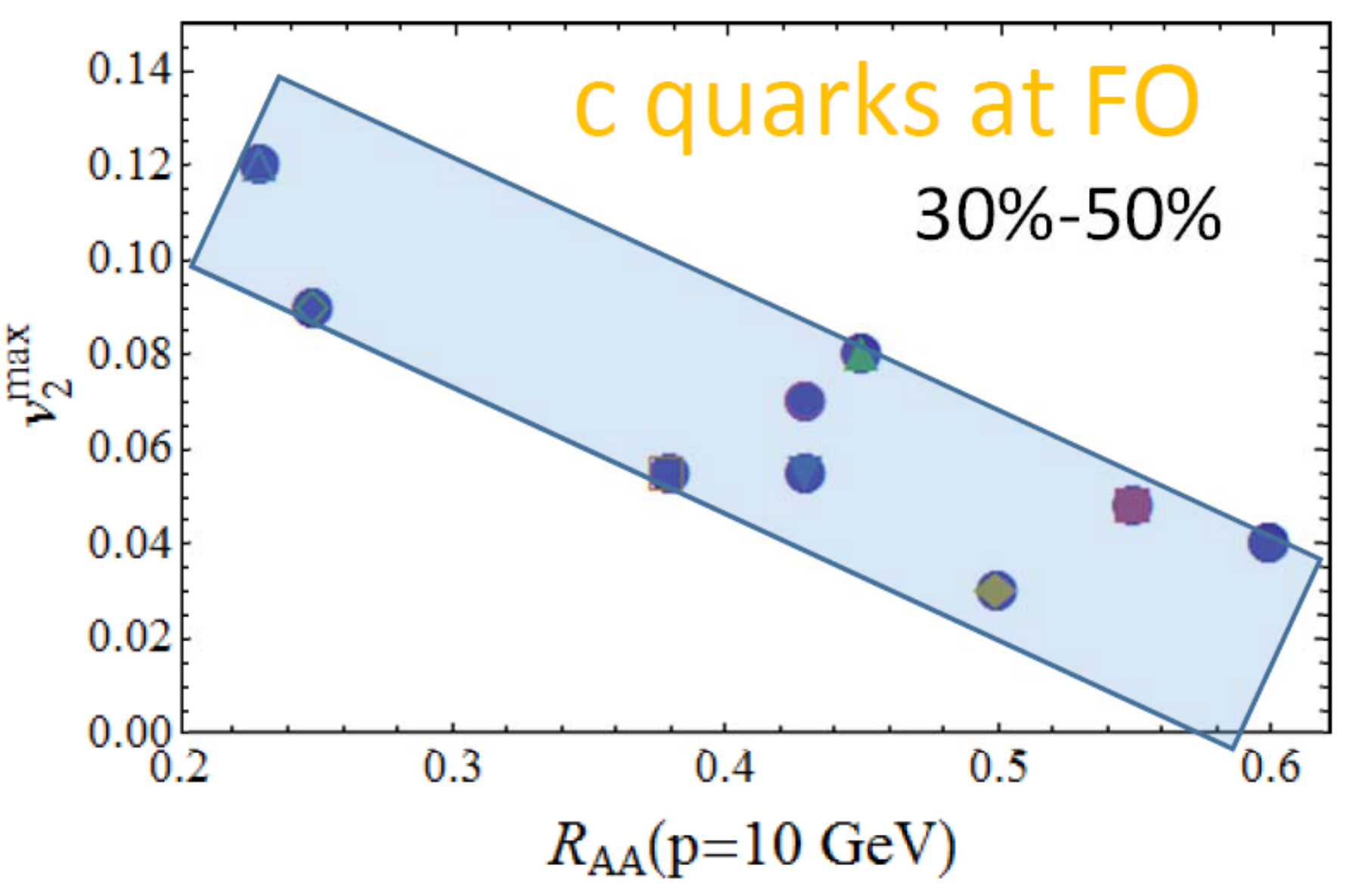} 
	\includegraphics[width=0.4\textwidth]{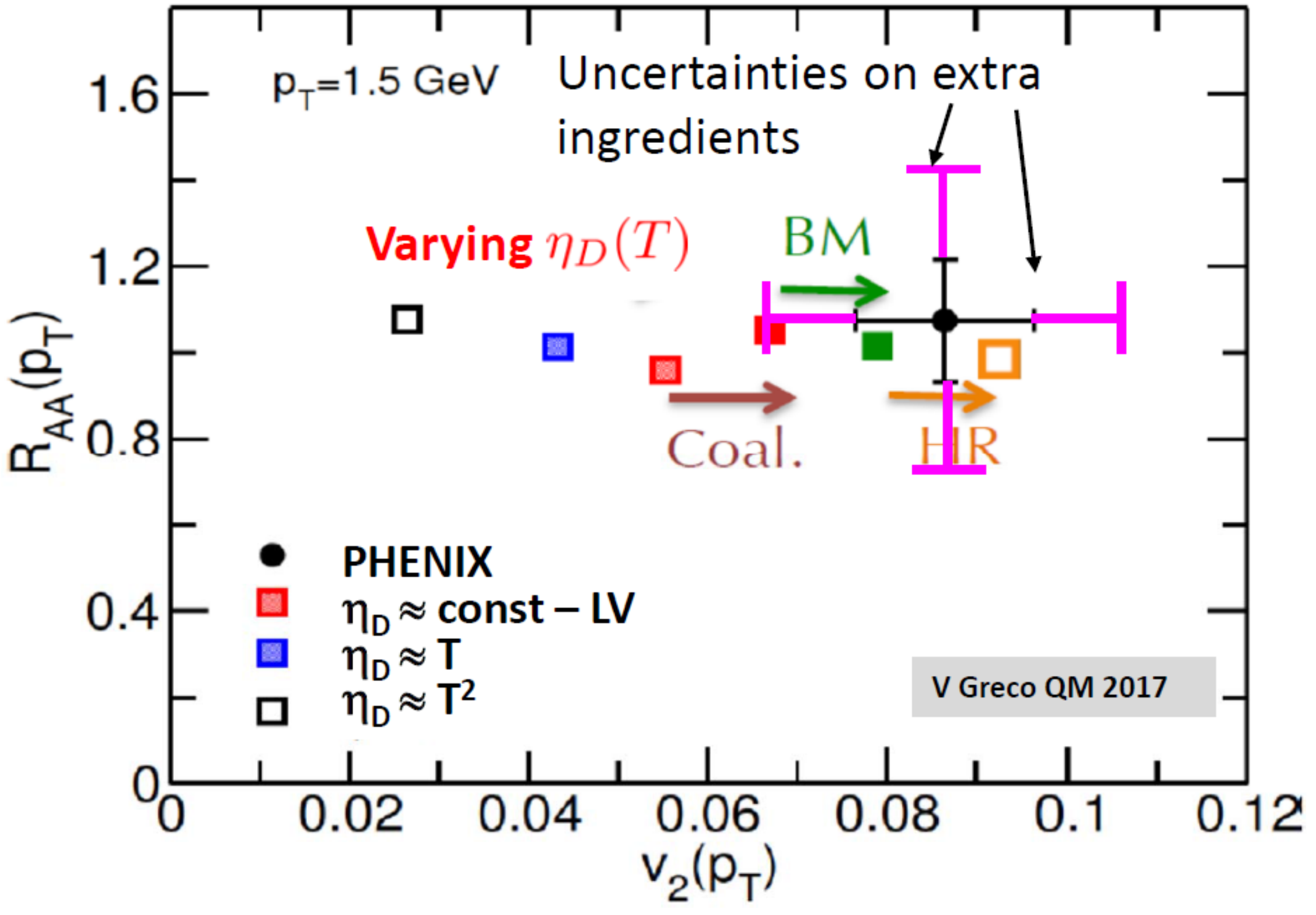} 
	\caption{Left: dots: correlation between ${\rm max}(v_2)$ and $R_{AA}(p_T=10\,{\rm GeV})$  for c-quarks at freeze out in various models compared in \cite{Rapp:2018qla} with the same energy loss; the band illustrates the "global" uncertainty. Right: accumulation of ingredients~\cite{Das:2015ana} leading to a correct agreement between $R_{AA}$ and $v_2$ with experimental error bars as well as theoretical uncertainties from the extra ingredients (basic figure from V. Greco's talk at QM 2017  ) }
	\label{fig:2}
\end{figure}

Since \cite{Andronic:2015wma}, several collective actions were undertaken in order to come up with strategies for extracting the FP coefficients, to estimate the influence of all "extra ingredients" in each model, as well as to suggest new strategies for model improvement based on deeper connection with fundamental theory. 
% cite Leiden.
In \cite{Rapp:2018qla}, several models (Catania, CUJET, Duke, TAMU, LBL-CCNU, MC@sHQ, pHSD, POWLANG, URQMD) were compared as for a) their initial c-quark spectra, b) several characteristics of the bulk, c) the transport implementation and d) the hadronization procedure. To test the effects of various bulk on HQ, it was decided to replace, in each model, the specific energy loss by a simple pQCD($\alpha_s=0.4$) energy loss cranked up by a factor 5. The $R_{AA}$ of c quarks at the end of the evolution was then found -- for most models -- to lay in a
0.3-0.4 (resp. 0.4-0.6) band for 0-10\% (resp 30\%-50\%) PbPb centrality class. Similar variations were observed for the $v_2(c)$ at freeze out. In fig.\ref{fig:2}, the correlation between ${\rm max}(v_2)$ and $R_{AA}(p_T=10\,{\rm GeV})$ for c-quarks at freeze out is illustrated. A valuable recommendation of \cite{Rapp:2018qla} in order to reduce uncertainties from the theory side is that each model should possibly adopt, besides its favorite one, a common state-of-the art bulk. This would at least permit to systematically estimate the uncertainty stemming from the bulk choice. While it is now admitted by most groups that HQ hadronization at the QGP freeze out should proceed through a dual process (interpolating between fragmentation at large $p_T$ and recombination at small $p_T$), significant deviations affect the various models, especially the recombination component, where 3 prescriptions coexist in the literature: the instantaneous parton coalescence (IPC), the resonance recombination model as well as the recently introduced in-medium fragmentation. While IPC is by far the most common choice,  
% cite refs
it should be stated that different parameters are often tuned to ensure that hadronization proceeds exclusively through recombination as $p_T \rightarrow 0$, with non trivial consequences at finite $p_T$. In \cite{Rapp:2018qla}, the uncertainties stemming from the hadronization where quantified resorting to the $H_{AA}$ quantity, defined as $H_{AA}=\frac{dN_D}{dp_T}/\frac{dN_{c,{\rm final}}}{dp_T}$, with deviations of the order of 25\% between various models. One concludes from \cite{Rapp:2018qla} that the combination of all sources of uncertainties for each model -- still for a common pQCD $\times 5$ energy loss --  is at least of the same order of magnitude as the experimental one, e.g. $\approx 15\%$ for $R_{AA}(D)$ and
$\approx 1\%$ for $v_2(D)$. On figure \ref{fig:2} (right), we illustrate the consequence of all uncertainties on the discriminating power of a $R_{AA}-v_2$ combined plot and conclude that better overall precision needs to be achieved in order to draw any firm conclusion.   

In \cite{Cao:2018} (convened by X.-N. Wang), an alternate approach was chosen in order to address the large discrepancies observed in the FP coefficients from various models: Adopting a uniform and constant medium (a so-called "brick problem"), the HQ interaction strength with the QGP was tuned in order to achieve an imposed value of the $R_{AA}$ at $p_T=15\,{\rm GeV/c}$ after an evolution lasting 3 fm/c. The "uncertainty band" for the tuned coefficients was then found to shrink accordingly, while clear structures emerged from this procedure, corresponding to various classes of interactions (quasi-particle model, pQCD-like, elastic + radiative). This study illustrates the maximal resolution that can be achieved for the extraction of FP coefficients if all extra ingredients would be taken the same way.

\section{"New" observables}
Recently, several observables were suggested which could further constrain the models and bring new insights on the FP coefficients, as f.i. the azimuthal correlations \cite{Nahrgang:2013saa} or momentum imbalance \cite{Uphoff:2013rka} between HF mesons. Although quite promising, such observables seem to be dominated by the initial state effects and to only differ by at most $\approx 10\%$ depending on the type of energy loss. They will thus require very good experimental accuracy. Two new observables were discussed at this conference: The directed flow $v_1$ of $D$ and $\bar{D}$ mesons as well as the 
$R_{AA}$ of $\Lambda_c$ baryons.$v_1(D/\bar D)$ was first suggested \cite{Das:2016cwd} to be generated by the initial strong magnetic field present in URHIC, with opposite signs for $D$ and $\bar{D}$ mesons, but it was then also argued~\cite{Chatterjee:2017ahy} to result from an initial tilt of the fireball in the reaction plane that generates such $v_1$ of the constituents independent of their charges, with same sign for $v_1(D)$ and $v_1(\bar D)$, in agreement with the trend presented by the STAR collaboration. Although it was mentioned that this observable could help in constraining the FP coefficients, the best strategy to me seems to first achieve a decent understanding of those coefficient and then use the $v_1(D/\bar D)$ to constrain extra ingredients like the initial magnetic field. Measuring $\Lambda_c$ is crucial as the parton-rich medium at FO could favor the recombination of HQ into baryons as compared to the p-p case, a conjecture supported by the recent measurements of the $\Lambda_c/D_0$ ratio at intermediate $p_T$ by the STAR and the ALICE experiments. Such observable could thus put further constrains on the recombination mechanisms which are the genuine footprint of reconfinement physics. While a state of the art coalescence scenario has been advanced this year \cite{Plumari:2017ntm}, some ambiguities subsist -- f.i. as for the prescription to be used for the normalisation -- and one has to be aware that allowing one free parameter for each new resonance drastically reduces the insights one can gain from considering several HF hadrons.    

\section{Conclusions and perspectives}
While in the past years, several approaches have been developed which make good contact with the experimental data, achieving robust conclusions as for the precise mechanisms and the interaction strength of HQ with the QGP will only be possible in the future by maintaining and even amplifying the recent collective efforts undertaken in order to provide more systematically errors from our models as well as “figures of merit” and adopt a reasonable base line for the "extra ingredients" which will make the role of the "core ingredients" more transparent. As discussed in  \cite{Aarts:2016hap,Rapp:2018qla}, making contact with the fundamental theory via lQCD is of paramount importance. In this respect, charm correlations and charm meson correlators  -- which indicate the presence of mesonic dof around $T_c$ -- should be evaluated in the existing and future models and compared with the lQCD results -- as performed f.i. in \cite{Liu:2017qah} -- as they allow to test the physical relevance of the various models. I have the conviction that this is a preliminary condition to reliably address the HF observables in small systems -- another fascinating topic not discussed in these proceedings -- where subtle initial and final state effects coexist and should thus be treated with great care.

%% The Appendices part is started with the command \appendix;
%% appendix sections are then done as normal sections
%% \appendix

%% \section{}
%% \label{}

%% References
%%
%% Following citation commands can be used in the body text:
%% Usage of \cite is as follows:
%%   \cite{key}         ==>>  [#]
%%   \cite[chap. 2]{key} ==>> [#, chap. 2]
%%

%% References with BibTeX database:

\bibliographystyle{elsarticle-num}
\bibliography{nuphaQMgossiaux}

\begin{thebibliography}{10}
\expandafter\ifx\csname url\endcsname\relax
  \def\url#1{\texttt{#1}}\fi
\expandafter\ifx\csname urlprefix\endcsname\relax\def\urlprefix{URL }\fi
\expandafter\ifx\csname href\endcsname\relax
  \def\href#1#2{#2} \def\path#1{#1}\fi

\bibitem{Liu:2017qah}
S.~Y.~F. Liu, R.~Rapp, {$T$-matrix Approach to Quark-Gluon Plasma}, Phys. Rev.
  C97~(3) (2018) 034918.
\newblock \href {http://arxiv.org/abs/1711.03282} {\path{arXiv:1711.03282}},
  \href {http://dx.doi.org/10.1103/PhysRevC.97.034918}
  {\path{doi:10.1103/PhysRevC.97.034918}}.

\bibitem{Andronic:2015wma}
A.~Andronic, et~al., {Heavy-flavour and quarkonium production in the LHC era:
  from proton–proton to heavy-ion collisions}, Eur. Phys. J. C76~(3) (2016)
  107.
\newblock \href {http://arxiv.org/abs/1506.03981} {\path{arXiv:1506.03981}},
  \href {http://dx.doi.org/10.1140/epjc/s10052-015-3819-5}
  {\path{doi:10.1140/epjc/s10052-015-3819-5}}.

\bibitem{Aarts:2016hap}
G.~Aarts, et~al., {Heavy-flavor production and medium properties in high-energy
  nuclear collisions - What next?}, Eur. Phys. J. A53~(5) (2017) 93.
\newblock \href {http://arxiv.org/abs/1612.08032} {\path{arXiv:1612.08032}},
  \href {http://dx.doi.org/10.1140/epja/i2017-12282-9}
  {\path{doi:10.1140/epja/i2017-12282-9}}.

\bibitem{Prino:2016cni}
F.~Prino, R.~Rapp, {Open Heavy Flavor in QCD Matter and in Nuclear Collisions},
  J. Phys. G 43~(9) (2016) 093002.
\newblock \href {http://arxiv.org/abs/1603.00529} {\path{arXiv:1603.00529}},
  \href {http://dx.doi.org/10.1088/0954-3899/43/9/093002}
  {\path{doi:10.1088/0954-3899/43/9/093002}}.

\bibitem{Rapp:2018qla}
A.~Beraudo, et~al., {Extraction of Heavy-Flavor Transport Coefficients in QCD
  Matter}\href {http://arxiv.org/abs/1803.03824} {\path{arXiv:1803.03824}}.

\bibitem{Abir:2015hta}
R.~Abir, A.~Majumder, {Drag-induced radiative energy loss from semihard heavy
  quarks}, Phys. Rev. C94~(5) (2016) 054902.
\newblock \href {http://arxiv.org/abs/1506.08648} {\path{arXiv:1506.08648}},
  \href {http://dx.doi.org/10.1103/PhysRevC.94.054902}
  {\path{doi:10.1103/PhysRevC.94.054902}}.

\bibitem{He:2015hfa}
L.~He, T.~Edmonds, Z.-W. Lin, F.~Liu, D.~Molnar, F.~Wang, {Anisotropic parton
  escape is the dominant source of azimuthal anisotropy in transport models},
  Phys. Lett. B753 (2016) 506--510.
\newblock \href {http://arxiv.org/abs/1502.05572} {\path{arXiv:1502.05572}},
  \href {http://dx.doi.org/10.1016/j.physletb.2015.12.051}
  {\path{doi:10.1016/j.physletb.2015.12.051}}.

\bibitem{Li:2018leh}
H.~Li, Z.-W. Lin, F.~Wang, {Charm quarks are more hydrodynamic than light
  quarks in elliptic flow}\href {http://arxiv.org/abs/1804.02681}
  {\path{arXiv:1804.02681}}.

\bibitem{Das:2015ana}
S.~K. Das, F.~Scardina, S.~Plumari, V.~Greco, {Toward a solution to the
  $R_{AA}$ and $v_2$ puzzle for heavy quarks}, Phys. Lett. B747 (2015)
  260--264.
\newblock \href {http://arxiv.org/abs/1502.03757} {\path{arXiv:1502.03757}},
  \href {http://dx.doi.org/10.1016/j.physletb.2015.06.003}
  {\path{doi:10.1016/j.physletb.2015.06.003}}.

\bibitem{Scardina:2017ipo}
F.~Scardina, S.~K. Das, V.~Minissale, S.~Plumari, V.~Greco, {Estimating the
  charm quark diffusion coefficient and thermalization time from D meson
  spectra at energies available at the BNL Relativistic Heavy Ion Collider and
  the CERN Large Hadron Collider}, Phys. Rev. C96~(4) (2017) 044905.
\newblock \href {http://arxiv.org/abs/1707.05452} {\path{arXiv:1707.05452}},
  \href {http://dx.doi.org/10.1103/PhysRevC.96.044905}
  {\path{doi:10.1103/PhysRevC.96.044905}}.

\bibitem{Xu:2017obm}
Y.~Xu, J.~E. Bernhard, S.~A. Bass, M.~Nahrgang, S.~Cao, {Data-driven analysis
  for the temperature and momentum dependence of the heavy-quark diffusion
  coefficient in relativistic heavy-ion collisions}, Phys. Rev. C97~(1) (2018)
  014907.
\newblock \href {http://arxiv.org/abs/1710.00807} {\path{arXiv:1710.00807}},
  \href {http://dx.doi.org/10.1103/PhysRevC.97.014907}
  {\path{doi:10.1103/PhysRevC.97.014907}}.

\bibitem{Cao:2016gvr}
S.~Cao, T.~Luo, G.-Y. Qin, X.-N. Wang, {Linearized Boltzmann transport model
  for jet propagation in the quark-gluon plasma: Heavy quark evolution}, Phys.
  Rev. C94~(1) (2016) 014909.
\newblock \href {http://arxiv.org/abs/1605.06447} {\path{arXiv:1605.06447}},
  \href {http://dx.doi.org/10.1103/PhysRevC.94.014909}
  {\path{doi:10.1103/PhysRevC.94.014909}}.

\bibitem{Li:2017wwc}
H.~T. Li, I.~Vitev, {Inverting the mass hierarchy of jet quenching effects with
  prompt $b$-jet substructure}\href {http://arxiv.org/abs/1801.00008}
  {\path{arXiv:1801.00008}}.

\bibitem{Zigic:2018ovr}
D.~Zigic, I.~Salom, M.~Djordjevic, M.~Djordjevic, {DREENA-B framework: first
  predictions of $R_{AA}$ and $v_2$ within dynamical energy loss formalism in
  evolving QCD medium}\href {http://arxiv.org/abs/1805.04786}
  {\path{arXiv:1805.04786}}.

\bibitem{Cao:2017crw}
S.~Cao, A.~Majumder, G.-Y. Qin, C.~Shen, {Drag Induced Radiation and
  Multi-Stage Effects in Heavy-Flavor Energy Loss}\href
  {http://arxiv.org/abs/1711.09053} {\path{arXiv:1711.09053}}.

\bibitem{Cao:2015hia}
S.~Cao, G.-Y. Qin, S.~A. Bass, {Energy loss, hadronization and hadronic
  interactions of heavy flavors in relativistic heavy-ion collisions}, Phys.
  Rev. C92~(2) (2015) 024907.
\newblock \href {http://arxiv.org/abs/1505.01413} {\path{arXiv:1505.01413}},
  \href {http://dx.doi.org/10.1103/PhysRevC.92.024907}
  {\path{doi:10.1103/PhysRevC.92.024907}}.

\bibitem{Kang:2016ofv}
Z.-B. Kang, F.~Ringer, I.~Vitev, {Effective field theory approach to open heavy
  flavor production in heavy-ion collisions}, JHEP 03 (2017) 146.
\newblock \href {http://arxiv.org/abs/1610.02043} {\path{arXiv:1610.02043}},
  \href {http://dx.doi.org/10.1007/JHEP03(2017)146}
  {\path{doi:10.1007/JHEP03(2017)146}}.

\bibitem{Nahrgang:2013saa}
M.~Nahrgang, J.~Aichelin, P.~B. Gossiaux, K.~Werner, {Azimuthal correlations of
  heavy quarks in Pb + Pb collisions at $\sqrt{s}=2.76$ TeV at the CERN Large
  Hadron Collider}, Phys. Rev. C90~(2) (2014) 024907.
\newblock \href {http://arxiv.org/abs/1305.3823} {\path{arXiv:1305.3823}},
  \href {http://dx.doi.org/10.1103/PhysRevC.90.024907}
  {\path{doi:10.1103/PhysRevC.90.024907}}.

\bibitem{Prado:2016szr}
C.~A.~G. Prado, J.~Noronha-Hostler, R.~Katz, A.~A.~P. Suaide, J.~Noronha, M.~G.
  Munhoz, M.~R. Cosentino, {Event-by-event correlations between soft hadrons
  and $D^0$ mesons in 5.02 TeV PbPb collisions at the CERN Large Hadron
  Collider}, Phys. Rev. C96~(6) (2017) 064903.
\newblock \href {http://arxiv.org/abs/1611.02965} {\path{arXiv:1611.02965}},
  \href {http://dx.doi.org/10.1103/PhysRevC.96.064903}
  {\path{doi:10.1103/PhysRevC.96.064903}}.

\bibitem{Gossiaux:2017zto}
P.~B. Gossiaux, J.~Aichelin, M.~Nahrgang, V.~Ozvenchuk, K.~Werner, {Coupled
  dynamics of heavy and light flavor flow harmonics from EPOSHQ}[EPJ Web
  Conf.171,18004(2018)].
\newblock \href {http://arxiv.org/abs/1711.01485} {\path{arXiv:1711.01485}},
  \href {http://dx.doi.org/10.1051/epjconf/201817118004}
  {\path{doi:10.1051/epjconf/201817118004}}.

\bibitem{Beraudo:2017gxw}
A.~Beraudo, A.~De~Pace, M.~Monteno, M.~Nardi, F.~Prino, {Development of
  heavy-flavour flow-harmonics in high-energy nuclear collisions}, JHEP 02
  (2018) 043.
\newblock \href {http://arxiv.org/abs/1712.00588} {\path{arXiv:1712.00588}},
  \href {http://dx.doi.org/10.1007/JHEP02(2018)043}
  {\path{doi:10.1007/JHEP02(2018)043}}.

\bibitem{Cao:2018}
S.~e.~a. Cao, {Towards extraction of heavy quark transport coefficients in
  quark-gluon plasma (in preparation)}.

\bibitem{Uphoff:2013rka}
J.~Uphoff, F.~Senzel, Z.~Xu, C.~Greiner, {Momentum imbalance of D mesons in
  ultra-relativistic heavy-ion collisions at LHC}, Phys. Rev. C89~(6) (2014)
  064906.
\newblock \href {http://arxiv.org/abs/1310.1340} {\path{arXiv:1310.1340}},
  \href {http://dx.doi.org/10.1103/PhysRevC.89.064906}
  {\path{doi:10.1103/PhysRevC.89.064906}}.

\bibitem{Das:2016cwd}
S.~K. Das, S.~Plumari, S.~Chatterjee, J.~Alam, F.~Scardina, V.~Greco, {Directed
  Flow of Charm Quarks as a Witness of the Initial Strong Magnetic Field in
  Ultra-Relativistic Heavy Ion Collisions}, Phys. Lett. B768 (2017) 260--264.
\newblock \href {http://arxiv.org/abs/1608.02231} {\path{arXiv:1608.02231}},
  \href {http://dx.doi.org/10.1016/j.physletb.2017.02.046}
  {\path{doi:10.1016/j.physletb.2017.02.046}}.

\bibitem{Chatterjee:2017ahy}
S.~Chatterjee, P.~Bożek, {Large directed flow of open charm mesons probes the
  three dimensional distribution of matter in heavy ion collisions}, Phys. Rev.
  Lett. 120~(19) (2018) 192301.
\newblock \href {http://arxiv.org/abs/1712.01189} {\path{arXiv:1712.01189}},
  \href {http://dx.doi.org/10.1103/PhysRevLett.120.192301}
  {\path{doi:10.1103/PhysRevLett.120.192301}}.

\bibitem{Plumari:2017ntm}
S.~Plumari, V.~Minissale, S.~K. Das, G.~Coci, V.~Greco, {Charmed Hadrons from
  Coalescence plus Fragmentation in relativistic nucleus-nucleus collisions at
  RHIC and LHC}, Eur. Phys. J. C78~(4) (2018) 348.
\newblock \href {http://arxiv.org/abs/1712.00730} {\path{arXiv:1712.00730}},
  \href {http://dx.doi.org/10.1140/epjc/s10052-018-5828-7}
  {\path{doi:10.1140/epjc/s10052-018-5828-7}}.

\end{thebibliography}

%% Authors are advised to use a BibTeX database file for their reference list.
%% The provided style file elsarticle-num.bst formats references in the required Procedia style

%% For references without a BibTeX database:

%\begin{thebibliography}{00}

%% \bibitem must have the following form:
%%   \bibitem{key}...
%%

% \bibitem{}

%\end{thebibliography}

\end{document}